\documentclass[aps,prl,preprint,groupedaddress]{revtex4-1}
\usepackage{amssymb}
\usepackage{amsmath}
\usepackage{amsbsy}
\usepackage[dvips]{graphics,epsfig}

\begin{document}



\title{Damped reaction field method and the accelerated convergence 
of the real space Ewald summation.}

\author{Victor H. Elvira and Luis G. MacDowell}
\email[]{lgmac@quim.ucm.es}
\affiliation{Departamento de Qu\'{\i}mica F\'{\i}sica, Facultad de Ciencias Qu\'{\i}micas,
Universidad Complutense de Madrid, 28040, Spain.}

\newcommand{\prom}[2]{\left \langle #1 \right \rangle_{#2}}
\newcommand{\du}{\mbox{$\,$} \mathrm{d}}
\newcommand{\Ppar}{p_{\parallel}}
\newcommand{\Pper}{p_{\perp}}
\newcommand{\derpar}[3]{
        \left(\frac{\partial #1}{\partial #2}\right)_{#3}
                      }
\newcommand{\Eq}[1]{Eq.~(\ref{eq:#1})}
\newcommand{\rvec}[2]{\mathbf{r}_{#1}^{#2}}
\newcommand{\kvec}[2]{\mathbf{k}_{#1}^{#2}}
\newcommand{\qvec}[2]{\mathbf{q}_{#1}^{#2}}
\newcommand{\tvec}[2]{\mathbf{t}_{#1}^{#2}}
\newcommand{\erfc}[1]{\mathrm erfc(#1)}


\date{\today}

\begin{abstract}
In this paper we study a general theoretical framework which allows to
approximate the real space Ewald sum by means of effective force shifted
screened potentials, together with a self term. Using this strategy it
is possible to generalize the reaction field method, as a means to
approximate the real space Ewald sum.  We show that this method
exhibits faster convergence of the Coulomb energy
than several schemes proposed recently in the literature while 
enjoying a much more sound and clear electrostatic 
significance. In terms of the damping parameter of the screened
potential, we are able to identify two clearly distinct regimes
of convergence. Firstly, a reaction field regime corresponding
to the limit of small screening, where effective pair potentials converge
faster than the Ewald sum. Secondly, an Ewald regime, where the plain
real space Ewald sum converges faster. Tuning the screening
parameter for optimal convergence occurs essentially at the
crossover. The implication is that effective pair potentials are
an alternative to the Ewald sum only in those cases where optimization
of the convergence error is not possible.
\end{abstract}

\pacs{}

\maketitle

\section{Introduction}

The increasing availability of large scale computer facilities is
allowing to study ever more complicated systems, with
greater detail as well as longer length and time scales.
Despite this progress, the accurate evaluation
of electrostatic interactions remains the most important bottleneck in 
molecular simulations of charged systems.\cite{allen86,frenkel02} 

This uncomfortable situation is reflected in the number of different
alternatives which are available in the literature in order to deal
with Coulombic 
interactions.\cite{barker73,deleeuw80,lekner89,hummer92,darden93,wolf99,tyagi05,shan05,fennell06}
Yet, it is clear that the benchmark for both efficiency and accuracy of
all studies remain the Ewald summation
technique.\cite{esselink95,toukmaji96,hummer99}

In this method, the full electrostatic energy of the system is split
into a real space contribution, which amounts to a pairwise summation
of an effective damped Coulomb potential, and a Fourier contribution,
which embodies the long range effects of the Coulomb interactions. The
latter term features a Fourier transform of the charge distribution,
which on the one hand, brings some conceptual
difficulties,\cite{deleeuw80,smith81,hummer99,ballenegger14} and on
the other, is very time consuming to calculate.\cite{kolafa92}

In the last decade, a number of studies have been devoted to
study more or less efficient methodologies that allow to calculate electrostatic
interactions while avoiding the cumbersome Fourier contributions
of the Ewald sum.\cite{wolf99,zahn02,fennell06,fukuda11,hansen12,fukuda13}
Such techniques, named recently under the provocative name of {\em pairwise
alternatives} to the Ewald sum,  have recently received  
considerable popularity, but also some degree of controversy
as regards efficiency,\cite{hansen12} and accuracy.\cite{mendoza08b,muscatello12}

The fact is that a pairwise alternative to the Ewald summation has
been available ever since the first few simulations of Coulombic
systems.\cite{barker73,neumann83,barker94} Indeed, the Reaction Field Method  is
almost as old as computer simulations of charged systems,\cite{allen86} 
yet, it has a clear theoretical background which more recent approaches lack
completely.\cite{zahn02,fennell06,hansen12} Despite this situation,
the Reaction Field Method seems to have been largely abandoned in
favor of other techniques, with some exceptions.\cite{miguez10}

Recently, Fukuda et al. proposed a heuristic approach to approximate
the Coulomb sum. This approach shares advantages of some of the
pairwise methods, in the sense that it screens the Coulomb interactions
with a fast decaying function, but has a somewhat more elaborate
electrostatic background.\cite{fukuda11} Indeed, it has been recently recognized
that this approach may be considered as a generalization of the reaction field 
method to screened potentials.\cite{kamiya13}

In this work we attempt to provide a sound
theoretical background for a generalized reaction field method that
achieves fast convergence of the Coulomb sum
as with several of the most popular pairwise
alternatives.\cite{wolf99,fennell06}
The theoretical calculations are supplemented with numerical
results for test systems, which demonstrate the superiority of
reaction field methods. Additionally, we perform detailed analysis
of the Coulomb sum convergence error  for either
reaction field and Ewald methods. This
will allow us to identify the region of convergence parameters
where each method is advantageous.

\section{Theoretical Background}

\subsection{Ewald Summation}

In most simulations of charged systems, the influence of
long range electrostatic interactions 
is estimated by assuming the finite sample is surrounded
by an infinite number of replicas. The energy felt
by, say, charge $i$, may be then estimated as a lattice
sum over the replicas:
\begin{equation}\label{eq:infsum}
  U_i = \sum_{{\bf n}={0}} \sum_j' \frac{q_iq_j}{|\rvec{ij}{}+{\bf n}|}
\end{equation} 
where ${\bf n}$ denotes a translation of the unit cell vector,
$q_i$ is the charge on particle $i$, and $\rvec{ij}{}$ is the position
of $j$ relative to $i$. Furthermore, it is understood that the
first sum runs over all possible unit cell translations, while the
second sum runs over all charges inside the unit box. A prime reminds
that $j$ must be different from $i$ when ${\bf n}={\bf 0}$.

This trick only gets rid of the boundary problem, but not of
the actual calculation of $U_i$, since the series is known to be
conditionally convergent, with a very slow convergence in the favorable 
cases. A particularly inconvenient case is a summation over spherical
shells, which is often not convergent.\cite{deleeuw80,wolf99}

In the classical treatment of Ewald, the conditionally convergent
lattice sum of charges is transformed into two rapidly convergent
series, such that:
\begin{equation}
  U_i = U^{\rm R}_i + U^{\rm F}_i  
\end{equation} 
with
\begin{equation}\label{eq:erfcsum}
      U^{\rm R}_i = 
  \sum_{{\bf n}={0}} \sum_j' 
 q_i q_j \frac{\erfc{\alpha |\rvec{ij}{}+{\bf n} |}}{ |\rvec{ij}{}+{\bf n} |}
\end{equation} 
and
\begin{equation}\label{eq:fourier}
      U^{\rm F}_i = 
  4\pi    \sum_{{\bf k}\ne {\bf 0}} \rho(\kvec{}{}) 
    \frac{ e^{-\frac{k^2}{4\alpha^2} } }{k^2} q_i e^{-i \kvec{}{} \cdot \rvec{i}{} } - q_i^2 \frac{\alpha}{\sqrt{\pi}}
\end{equation}
where $\kvec{}{}$ are vectors in Fourier space
and  $\rho(\kvec{}{})$ is the Fourier transform of the charge density.
Additionally, the Ewald sum may contain a surface term, $U_{\rm surf}$, which
accounts for the boundary conditions of the system. Such term arises strictly
from the long range  electrostatic interactions with the boundary and
surrounding medium, and is therefore
 essentially  a Fourier contribution corresponding to the missing $k=0$ term
of the reciprocal space sum.\cite{deleeuw80,smith81,ballenegger14} In practice,
for bulk systems under metallic boundary conditions the surface term may
be neglected, so that we will henceforth drop this complication.

The inverse length, $\alpha$, plays a key role, dictating the
convergence of the series. A large value of $\alpha$ leads
to a rapidly convergent real space series that can be actually
truncated already at ${\bf n}=0$, but then the $U^{\rm F}$
contributions converges slowly. Alternatively, a small value of $\alpha$
produces a very fast convergence of $U^{\rm F}$, but then
$U^{\rm R}$ is slowly convergent. 

An important observation made by Wolf et al. is that for moderately
small values of $\alpha$, most of the Fourier contribution is actually
given by the simple self term, 
\begin{equation}\label{eq:selfew}
U^{\rm self}=-q_i^2 \frac{\alpha}{\sqrt{\pi}}, 
\end{equation} 
so that
the expensive reciprocal space summation may be ignored
altogether.\cite{wolf99}

This observation has allowed for the recent development of efficient
methodologies for the calculation of electrostatic 
interactions.\cite{wolf99,zahn02,fennell06,fukuda11}

\subsection{Accelerating the convergence of the real space sum}

Having circumvented the problem of calculating the expensive reciprocal space
sum, there still remains a crucial issue:
how fast is the convergence of 
$U^{\rm R}$ in those cases where the reciprocal space summation
may be ignored? 

For most practical purposes, $\erfc{\alpha r}$ decays so fast that
the lattice summation required for the evaluation of $U_i^{\rm R}$ may be 
ignored. Rather, a plain spherical cutoff
is usually employed for distances larger than a cutoff radius, $R_c$, with
the hope that terms of order $\erfc{\alpha R_c}$ or smaller may be
neglected. However, it remains desirable to have a cutoff as small as
possible. For this purpose, one may try to exploit the techniques of
continuum electro--dynamics for the $\erfc{\alpha r}/r$ potential in
the same spirit as it has been successfully done for the plain Coulomb
potential.\cite{barker69,neumann83,hummer92,wolf99} This would allow
to truncate $U_i^{\rm R}$ at cutoffs where $\erfc{\alpha r}$ is
actually not negligible.\cite{wolf99,zahn02,fennell06,fukuda11}

\begin{figure}
\begin{tabular}{ccc|ccc}
\includegraphics[width=0.20\textwidth]{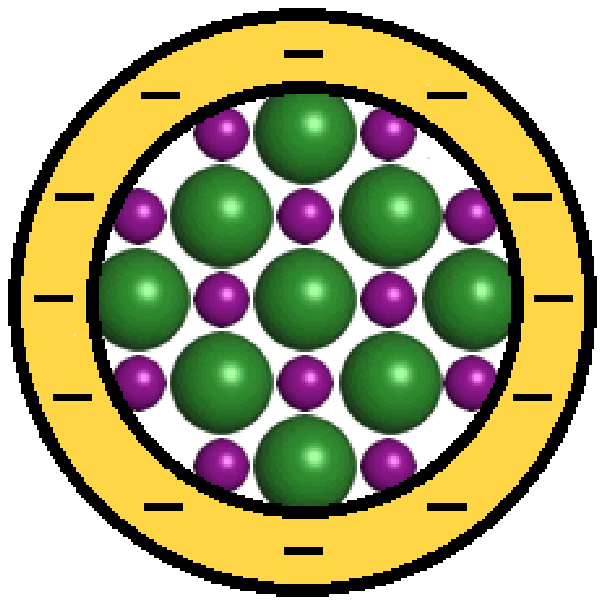} &
\raisebox{1.5cm}{$\leftarrow$} &
\includegraphics[width=0.20\textwidth]{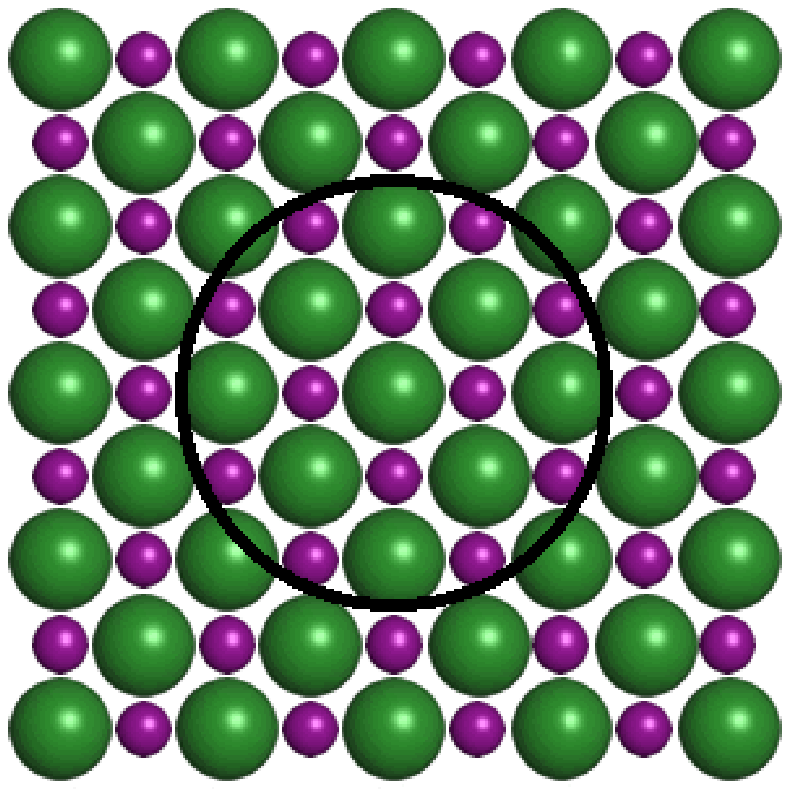} &
\includegraphics[width=0.20\textwidth]{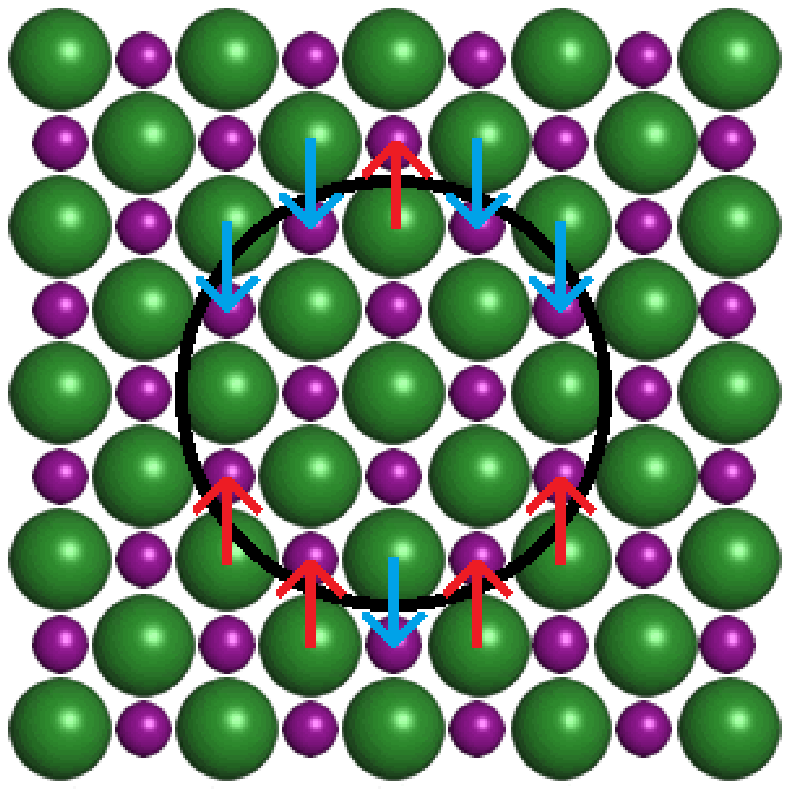} &
\raisebox{1.5cm}{$\rightarrow$} &
\includegraphics[width=0.20\textwidth]{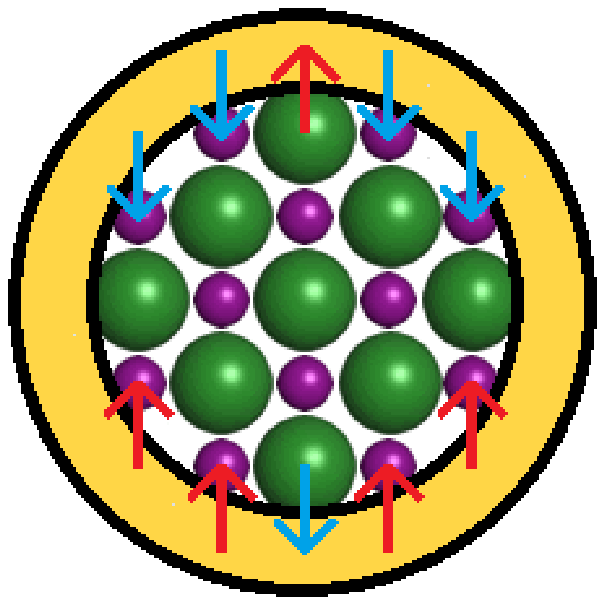} 
\end{tabular}
\caption{\label{fig:scheme} Sketch of the charge neutralization
schemes employed in the Wolf  and Damped Reaction Field
methods. In the Wolf method (left), the net charge
inside the sphere surrounding the central ion is uniformly 
spread over its surface. In the Damped Reaction Field Method (right),
an additional surface charge due to broken effective dipoles
between the central ion and the surrounding charges is considered. 
For a perfectly regular crystal lattice, the
charge distribution is also uniform, but will usually be
non-uniform and therefore produce a net force.
}
\end{figure}

In this regard, another important observation made by Wolf is
that Coulomb sums over spherical shells show good convergence
whenever the net charge inside the sphere vanishes (Fig. 1).\cite{wolf92} Whence,
the poor convergence is due to the the fact that spheres
of arbitrary radius centered about an ion $i$ will usually carry a net charge, 
$\Delta q_i$. The convergence of the Coulomb sum over spheres
can be much improved by summing over charge neutralized spheres.
This may be achieved in practice by assuming a fictitious 
uniform charge $\sigma_i^{\rm charge}$  over the surface of the 
sphere:\cite{wolf99}
\begin{equation}\label{eq:usig}
   \sigma_i^{\rm charge} = \frac{1}{4\pi R_c^2 } \sum_{j}^* q_j 
\end{equation} 
where the asterisk superscript indicates that the summation is
restricted to those particles $j$ inside the cutoff sphere (such that
$r_{ij}<R_c$).

Whence, for a source charge $i$ with arbitrary electric potential $q_i\phi(r)$, an
improved approximation for the energy is:
\begin{equation}
 U_i \approx  \sum_{j}^*
         q_i q_j\phi(r_{ij}) -  q_i\Delta q_i\phi(R_c)
\end{equation} 
Surprisingly, this charge neutralization may be implemented by means of an
effective pairwise potential:\cite{wolf99}
\begin{equation}\label{eq:sts}
 u =  q_i q_j  \{ \phi(r_{ij}) - \phi(R_c)   \}
\end{equation} 
where the first term in the right hand side accounts for the bare interaction
from the potential $q_i\phi(r)$, while the second term is the energy
resulting from the interaction with the fictitious charge distribution of
\Eq{usig}.
Clearly, such interaction merely shifts the pair potential.
Therefore, it does not result in an additional force. 

However, one expects that the truncation of
interactions at $R_c$ will not only produce a spurious net charge about ion $i$,
but also a net polarization (Fig. 1).\cite{wolf92} The force stemming from the uniform net charge is
zero for reasons of symmetry, but  the net polarization will
result in a finite electric field on charge $i$.\cite{fukuda11}

This idea  may be elaborated quantitatively in terms of the
sphere's polarization and the laws of electrostatics, which dictate
that a polarized dielectric produces a surface charge density
of magnitude:
\begin{equation}\label{eq:surfcharge}
  \sigma_i^{\rm pol}({\bf e}) = {\bf P_i} \cdot {\bf e}
\end{equation} 
where ${\bf P}_i$ is the net polarization inside the sphere of radius $R_c$ 
centered on $i$ and ${\bf e}$ is a unit vector normal to the surface.
Since the medium is overall neutral, we assume that the charges
beyond the sphere act such as to cancel exactly this charge distribution.
Accordingly, the full charge neutralizing distribution is:
\begin{equation}
   \sigma_i^{\rm neutral} = - ( \sigma_i^{\rm charge} + 
         \sigma_i^{\rm pol}({\bf e}) )
\end{equation} 

The particle $i$ bears a potential $\phi(r)$, and thus exerts
a force $-\nabla \phi(r)$ on whichever other charge.
Now, consider the infinitesimal force, $\du {\bf F}_i^{\rm surf}$ felt by
charge $i$ due to the charge neutralizing distribution on an infinitesimal 
surface element $\du S = R^2 \sin\theta\du\theta\du\phi$: 
\begin{equation}
  \du {\bf F}_i^{\rm surf} = - q_i \left . \nabla \phi\;\right |_{R_c} \;
               \sigma_i^{\rm neutral} \du S
\end{equation} 
Integrating over all the sphere's surface, the term
stemming from the uniform distribution $\sigma_i^{\rm charge}$ vanishes for
reasons of symmetry, but the non uniform $\sigma_i^{\rm pol}({\bf e})$ term 
yields:
\begin{equation}
  {\bf F}_i^{\rm surf} = \frac{4\pi}{3} q_i 
         \left . \frac{\du \phi}{\du r}\; \right |_{R_c} R_c^2 \; {\bf P}_i
\end{equation} 
After substitution of ${\bf P_i}$ in terms of the explicit charges inside
the sphere we obtain:
\begin{equation}
    {\bf F}_i^{\rm surf} = q_i \sum_{j}^* q_j 
  \left . \frac{\du \phi}{\du r}\;\right |_{R_c}
        \frac{r_{ij}}{R_c} \frac{\rvec{ij}{}}{r_{ij}}
\end{equation} 
The total force effectively felt on $i$ then contains the actual
interactions between particles inside the sphere, together with the
surface term accounting effectively for interactions with the
remaining charges:
\begin{equation}
  {\bf F}_i = - \sum_{j}^* q_i q_j \left . \nabla \phi \right |_{r_{ij}}
         + q_i \sum_{j}^* q_j \left . \frac{\du \phi}{\du r} \right |_{R_c}
     \frac{r_{ij}}{R_c} \frac{\rvec{ij}{}}{r_{ij}}
\end{equation} 
Clearly, the net force may be cast in terms of an effective pair potential
of the form:
\begin{equation}\label{eq:gf}
  {\bf F}_{ij} =  - q_i q_j \left (
                     \left .  \frac{\du \phi}{\du r} \right |_{r_{ij}} - 
             \left .   \frac{\du \phi}{\du r} \right |_{R_c}
               \frac{r_{ij}}{R_c} 
                    \right ) \frac{\rvec{ij}{}}{r_{ij}}
\end{equation} 
This is the formal result we sought for. It corresponds to a continuous
linearly screened force which smoothly vanishes at the cutoff.

\subsection{Discussion}

In order to understand its  significance, it is convenient to recall 
the result for the effective electrostatic potential of the Reaction 
Field Method, which assumes a continuous dielectric medium with dielectric 
constant $\epsilon$ beyond the cutoff:\cite{neumann85}
\begin{equation}\label{eq:rf85}
  {\bf F}_{ij} =   q_i q_j \left (
             \frac{1}{r_{ij}^2} - 
          \frac{2\epsilon-2}{2\epsilon+1}\frac{1}{R_c^2}\frac{r_{ij}}{R_c} 
                    \right ) \frac{\rvec{ij}{}}{r_{ij}}
\end{equation} 
Comparing this result with \Eq{gf}, it is clearly seen that our
result is recovered for the special case where $\phi(r)=1/r$, as in
the Coulomb potential, and the additional assumption of $\epsilon=\infty$.
{\em Whence,  \Eq{gf} corresponds to a generalization of the Reaction Field Method for arbitrary electric potentials, and a particular choice for
the boundary conditions at the surface of the cutoff sphere}. The
Reaction Field Method  shows that taking into account explicitly a finite 
susceptibility beyond the cutoff requires to assume a surface charge density 
of $\sigma_i^{\rm pol}=(2\epsilon-2)/(2\epsilon+1) {\bf P_i}\cdot {\bf e}$
in place of \Eq{surfcharge}.

On the other hand, the assumption embodied in \Eq{surfcharge}, and
other methods,\cite{wolf99,fukuda11,fukuda13}
implies that the charges left out beyond the cutoff are fully available
to screen the net dipole created inside the sphere. This statement
may be understood if we consider yet another refinement over the
Reaction Field Method, namely, to assume that the charge distribution
outside the cutoff sphere is given as  a Boltzmann weighted average
dictated by the field of the ions inside the sphere. This task
may be accomplished in the Debye-H\"uckel approximation, and
yields a generalization of the Reaction Field Method that accounts
for interactions with both a dielectric continuum and a smooth
charge distribution of given concentration:\cite{barker94,tironi95}
\begin{equation}\label{eq:rfdh}
  {\bf F}_{ij} =   q_i q_j \left (
             \frac{1}{r_{ij}^2} - 
          \frac{(2\epsilon-2)(1+\kappa R_c)+\epsilon(\kappa R_c)^2}
               {(2\epsilon+1)(1+\kappa R_c)+\epsilon(\kappa R_c)^2}
          \frac{1}{R_c^2}\frac{r_{ij}}{R_c} 
                    \right ) \frac{\rvec{ij}{}}{r_{ij}}
\end{equation} 
where $\kappa$ is the inverse Debye screening length of the
free charges. For vanishing concentration of charges, $\kappa=0$, and
this model recovers \Eq{rf85} exactly. In the opposite limit, $\kappa$ becomes
infinite, and then \Eq{rfdh} becomes equal to the Reaction Field Method with
conducting boundary conditions. This illustrates our statement, that the
approximation \Eq{surfcharge} corresponds to an infinite availability of
charges for screening of the dipole inside the cutoff sphere.
Unfortunately, the accuracy of Debye-H\"uckel theory  is limited to
very low ion concentration, so the refinement of \Eq{rfdh} is more a 
conceptual improvement than an accurate working equation at typical
simulation conditions.

The fact is, once the interactions are truncated beyond a cutoff, one
can not do without an arbitrary approximation as to the
charge distribution of the surrounding medium. The first obvious
choice is to assume a dielectric response, but then the precise
value of the dielectric constant needs to be specified.
In simulations of molten salts or ionic
fluids, it  seems reasonable to choose metallic boundary conditions.
On the other hand, in simulations of polar fluids 
the choice of $\epsilon$ equal to the fluid's dielectric constant  
is more natural. If, however, the polar fluid is simulated with
explicit charges, as is usually the case, there is then not an
obvious choice. 

Fortunately, most solvents of interest in studies of charged systems,
and particularly water, have a rather large dielectric constant, so
that the ratio $(2\epsilon-2)/(2\epsilon + 1)$ is very close to unity,
as in the case of $\epsilon=\infty$. Furthermore, for finite $\epsilon$,
the force of the RFM becomes discontinuous at the cutoff, and this severely
hampers applications in Molecular Dynamics. Such inconvenience  may
be altogether avoided, since  it has
been shown  that the precise choice of dielectric boundary conditions 
does not significantly change the outcome of the simulations, particularly for
the case of phase coexistence.\cite{garzon94,miguez10} 

For these reasons, we believe the boundary conditions that are implied in 
\Eq{gf} are the most judicious choice for condensed
phases of both i) polar fluids with high 
dielectric constant and ii) molten salts that will be studied in the next 
section. Indeed, they correspond to the accepted choice in a large body of
simulations.\cite{garzon94,wolf99,miguez10,fukuda11,fukuda13} 
A word of caution is required for applications to very low density systems.
In such cases, where typically ions are separated by large distances
and the Debye screening length is very large,
one cannot expect that the net charge about a single isolated ion
will be neutralized  at all within a small cutoff. In such cases, an Ewald
type summation might be the only reliable alternative.

\subsection{Damped force}

The advantage of \Eq{gf} over the traditional RFM is that it allows
to accelerate the convergence by using a fast decaying electric
field $\phi(r)$ in place of the Coulomb potential. 

For $\phi(r) = \erfc{\alpha}/r$, it yields a generalized
damped reaction field equation for the force, as:
\begin{equation}\label{eq:drff}
   F_{ij} = q_i q_j  \left \{ \left [ 
   \frac{ \erfc{\alpha r_{ij}} }{ r_{ij}^2 }
 + \frac{ 2\alpha }{ \sqrt{\pi} } \frac{ e^{-\alpha^2 r_{ij}^2} }{r_{ij}} 
                             \right ]
 -                           \left [ 
  \frac{\erfc{\alpha R_c}}{R_c^2} + \frac{2\alpha}{\sqrt{\pi}} 
  \frac{e^{-\alpha^2 R_c^2}}{R_c} 
                             \right ]
        \frac{r_{ij}}{R_c}
                     \right \}
\end{equation} 
For the special case where $\alpha=0$, we recover the result 
of \Eq{rf85} with $\epsilon=\infty$.

It is difficult here to know exactly how \Eq{drff} is related
to the corresponding expressions given by Wolf et al.\cite{wolf99}
The reason is that these authors obtain their forces unconventionally
as an imaginary process where $r_{ij}\to R_c$ in order to enforce their 
potential to yield a shifted force, which
would otherwise not be the case. Whence the authors write:
\begin{equation}
  F_{ij} = q_i q_j  \left \{ \frac{1}{r_{ij}^2} - 
          \frac{1}{R_c^2} \left . \frac{r_{ij}}{R_c} \right |_{r_{ij}=R_c} 
                    \right \}
\end{equation} 
We find difficult to interpret what the condition $r_{ij}=R_c$  means.
If we simply ignore the odd condition, we recover \Eq{gf}. 
If, as 
interpreted by Fennell and Gezelter, we take the equality as
written, we then obtain merely a shifted force potential of
little  electrodynamic significance.\cite{fennell06}

\subsection{Formal treatment}

In the previous section we have obtained our results using elementary
electrodynamics and back of the envelope arguments. Here we 
derive our results using a more formal treatment of molecular
electrodynamics as formulated by Neumann, Boresch
and Steinhauser.\cite{neumann83,boresch01}

Our derivation starts with a basic equation for the electric
field that results from the polarization ${\bf E}({\bf r})$ of a
uniform medium due to an external source field, ${\bf E_0}$:
\begin{equation}
 {\bf E}(\rvec{}{}) = {\bf E_0}(\rvec{}{}) + 
       \int \du \rvec{}{'} 
      {\bf T}(\rvec{}{}-\rvec{}{'}) \cdot {\bf P}(\rvec{}{'})
\end{equation} 
where ${\bf T}(\rvec{}{})=\nabla \nabla \phi$ is the dipole--dipole
tensor, and the integration is over the simulation cell under
toroidal boundary conditions. This equation was instrumental in
the establishment of a consistent framework for the calculation
of dielectric relaxation phenomena in the 1980's,\cite{neumann83,neumann84}
and has been exploited recently to study the dielectric constant
of anisotropic media.\cite{macdowell10}

An important point here is the realization that this equation is
also valid for the case where ${\bf E_0}(\rvec{}{})$ results from
atomic charge distributions interacting via a generalized
Green's function $\phi(\rvec{}{})$,
whether it is a Coulomb interaction  or some other modified
central force potential. The only caution is to 
keep in mind that the dipole--dipole tensor has to be accordingly
modified as in the definition above. Taking these two considerations
into account, we can write an equation for the electric field
on charge $\rvec{i}{}$ that results from the presence of a second
charge $\rvec{j}{}$ interacting with $i$ via the generalized
green function $\phi_{\rm mod}(r) = S(r)/r$, with $S(r)$ an arbitrary
function:\cite{boresch01}
\begin{equation}\label{eq:dscft}
 {\bf E}(\rvec{i}{}) = -\nabla \phi_{\rm mod}(r_{ij}) + 
       \int \du \rvec{}{} 
      {\bf T}_{\rm mod}(\rvec{i}{}-\rvec{}{}) \cdot {\bf P}(\rvec{}{})
\end{equation} 
where now ${\bf T}_{\rm mod}(\rvec{i}{})=\nabla \nabla \phi_{\rm mod}(r)$.
For our purposes, it proves convenient to express ${\bf T}_{\rm mod}$
in terms of the dipole--dipole tensor corresponding to a Coulomb 
potential:\cite{boresch01}
\begin{equation}
  {\bf T}_{\rm mod} = ( S - rS' + \frac{1}{3}r^2 S'' ){\bf T} +
               \frac{1}{3} \frac{S''}{r}{\bf I}
\end{equation}
where ${\bf I}$ is a unit matrix and the prime indicates derivation
with respect to $r$.

In order to perform the integral of \Eq{dscft}, one needs to take into
account that ${\bf T}(\rvec{}{})$ is an odd function, except at the singularity
$\rvec{}{}\to{\bf 0}$, where the tensor is:\cite{neumann83}
\begin{equation}
  \lim_{r\to 0} {\bf T}(\rvec{}{}) = -\frac{4\pi}{3} \delta(\rvec{}{})
\end{equation} 
As a result, the integral does not vanish altogether, but rather,
yields:
\begin{equation}
 {\bf E}(\rvec{i}{}) = -\nabla \phi_{\rm mod}(r_{ij}) -
 \frac{4\pi}{3} S(0) \; {\bf P}(\rvec{i}{})
+ \frac{1}{3} \int\du \rvec{}{} 
       \frac{S''}{r} \; {\bf I}\cdot {\bf P}(\rvec{}{})
\end{equation} 
this is a general result for a wide choice of $S$ functions,
including whatever polynomial, the exponential or the complementary error 
function. The integral that remains is difficult to solve for
the general case of a position dependent polarization. However,
assuming constant polarization within the cutoff sphere, the
integral can be solved by parts, yielding:
\begin{equation}
  {\bf E}(\rvec{i}{}) = -\nabla \phi_{\rm mod}(r_{ij}) -
 \frac{4\pi}{3} \phi_{\rm mod}'(R_c) R_c^2\; {\bf P}_i
\end{equation} 
where ${\bf P}_i$ is the uniform polarization inside the sphere
about $i$, that is, ${\bf P}_i = q_j \rvec{ij}{}/4\pi R_c^3$.
Substitution of ${\bf P}_i$ into the above result and
multiplying by the charge $q_i$ at $i$ then leads
right away to the general result obtained in the previous
section, \Eq{gf}.

The significance of calculating an electric field rather than
a potential can be now understood, since  
a continuous potential at $r=R_c$ results ab--initio
by integration of the force:
\begin{equation}
    u(r) = - \int_{\infty}^{r} F(r) \du r
\end{equation} 
considering that $F(r)$ vanishes beyond $R_c$, we obtain:
\begin{equation}\label{eq:gv}
  u(r_{ij}) = q_iq_j \left \{ \phi_{\rm mod}(r_{ij}) - \phi_{\rm mod}(R_c) - \frac{1}{2} 
            \phi_{\rm mod}'(R_c) \; \frac{ r_{ij}^2 - R_c^2}{R_c}
                     \right \}
\end{equation} 
Clearly, the resulting potential not only shifts $\phi_{\rm mod}(r)$
but also produces an additional polarization force (c.f. \Eq{sts}).
The equation has the form of a generalized reaction field
potential, that may be employed to improve the summation of
whatever generalized $\phi(r)$ function. For $\phi=1/r$ we
recover the reaction field model of Neumann,\cite{neumann83}
properly modified to produce a continuous force at $r=R_c$
as in Ref.\onlinecite{hummer92}. This is an advantage over
other treatments, where the continuous form is
implemented add--hoc on the basis of numerical convenience 
(see DL-POLY or Gromacs reference manuals).

In order to complete the formulation, we further need to
supplement our model with a self term that gauges the
total energy, accounting for the difference between the
actual and modified potentials:\cite{boresch01}
\begin{equation}\label{eq:ug}
  U_i^{\rm self} = \lim_{r\to 0 }\quad \frac{1}{2} \; q_i^2 
       \left ( u(r) - \phi_{\rm mod}(r) \right )
\end{equation} 
Using \Eq{gv}, we obtain the general expression for the self term
of the generalized reaction field method.
\begin{equation}\label{eq:uselfg}
  U_i^{\rm self} = - \frac{1}{2} \; q_i ^2 \left ( \phi_{\rm mod}(R_c) - 
      \frac{1}{2} \; \phi'_{\rm mod}(R_c) R_c
            \right )
\end{equation} 

\Eq{gv} and \Eq{uselfg} are the more important results of this section.
For the special case where $\phi(r)=1/r$, we recover exactly the
reaction field model of Hummer et al.\cite{hummer92} 
Here, we show that the  reaction field form
is kept generally for whatever Green function $\phi_{\rm mod}(r)$. As a
result, it may be exploited also to {\em accelerate the convergence}
of damped coulomb potentials in Ewald type summations.

\subsection{Pairwise schemes for the calculation of the real space Ewald
summation}

We finish this section with the pair potential and self terms
that result for probably the most significant damped coulomb
potential, namely, $\phi_{\rm mod}(r)=\erfc{\alpha r}/r$.
Using \Eq{gv} and \Eq{uselfg}, we obtain:\cite{elvira12}
\begin{equation}\label{eq:udrf}
   u_{\rm ercf}(r_{ij}) = q_iq_j \left \{
   \frac{\erfc{\alpha r_{ij}}}{r_{ij}} - \frac{\erfc{\alpha R_c}}{R_c}
   + \frac{1}{2} \left [  \frac{\erfc{\alpha R_c}}{R_c^2} 
     + \frac{2\alpha}{\sqrt{\pi}} \frac{e^{-\alpha^2 R_c^2}}{R_c}
                 \right ] \frac{r_{ij}^2 - R_c^2}{R_c}
    \right \}
\end{equation} 
While for the self term, we have:
\begin{equation}\label{eq:uselfdrf}
  U_i^{\rm self} =  -\frac{1}{2} q_i^2
 \left \{ \frac{\erfc{\alpha R_c}}{R_c} + 
     \frac{1}{2} \left ( 
  \frac{\erfc{\alpha R_c}}{R_c} + 
  \frac{2\alpha}{\sqrt{\pi}} e^{-\alpha^2 R_c^2}
     \right )
 \right \}
\end{equation} 
Notice that this self energy corrects for the truncation
of  the exact real space lattice summation of $\phi_{\rm mod}(r)=\erfc{\alpha r}/r$,
\Eq{erfcsum}. 
i.e., it cannot possibly account for the Fourier space term of  \Eq{fourier}. 
If the purpose is to approximate the full Coulomb sum, \Eq{infsum}, 
and $\alpha$ is chosen such that the reciprocal space sum may be ignored,
one must still account for the self term of the Fourier sum, \Eq{selfew}.

Actually, this point can be illustrated from the formalism afforded by 
Eq.~\ref{eq:dscft}, \ref{eq:gv}, \ref{eq:ug}. To see this,
consider the situation where
this approach is applied to approximate the infinite sum  \Eq{infsum}
of coulomb potentials, $1/r$, instead of the related $\erfc{r}/r$ sum
of \Eq{erfcsum}. 
Rather than employing \Eq{uselfg} for the self term, we
would then need to consider the self term as the limiting value
of  $ -\frac{1}{2} q_i^2 ( u(r) - 1/r )$, since now the excess energy
is over $\phi=1/r$, rather than over $\phi_{\rm mod}=\erfc{r}/r$. 
The self term would still be  as \Eq{uselfdrf}, plus an
extra term $\lim\, r\to 0$ of $-\frac{1}{2}\, q_i^2\, {\mathrm erf}(r)/r$. One
can recognize here exactly the self term of the Fourier contribution,
\Eq{selfew}.  Therefore, the theoretical treatment provides naturally
the generalized reaction field result of 
\Eq{udrf} and \Eq{uselfdrf}, 
plus the Ewald sum self term as the best possible approximation to
\Eq{infsum} that can be obtained by performing
 a truncated sum of $\erfc{r}/r$ terms.

The general results \Eq{gv}, \Eq{uselfg}, as well as \Eq{udrf}--\Eq{uselfdrf} for the damped Coulomb
potential are exactly as the Zero Charge-Zero Dipole method derived  recently
by Fukuda et al. using a heuristic approach based on the mirror image technique.\cite{fukuda11} Our
results provide an alternative derivation  showing  clearly the strong
connection with the Reaction Field technique.\cite{hummer92} As such, we will henceforth
refer to \Eq{udrf}--\eqref{eq:uselfdrf} as  Damped Reaction Field (DRF) Method.

The results \Eq{udrf}--\eqref{eq:uselfdrf} resemble effective pair potentials
that have been suggested
recently.\cite{wolf99,zahn02,fennell06,fukuda11,fukuda13}
Indeed, they include a shift in the potential that
is at the heart of Wolf's method, but provide 
also an additional force as in the Fennel and Gezelter method and 
related approaches.\cite{fennell06,fukuda11,fukuda13} 

In fact, all of these methods may be considered as a wide class, that,
by use of Eq.~\ref{eq:dscft}, \ref{eq:gv}, \ref{eq:ug}, allows to 
approximate the real space 
Ewald summation of \Eq{erfcsum}.  This is achieved using an effective pairwise
potential of the form:
\begin{equation}\label{eq:pairwisef}
 u(r_{ij}) = q_i q_j \left \{
 \begin{array}{cc}
 f(r_{ij};\alpha,R_c) & \quad r_{ij} \le R_c\\
  0 & \quad r_{ij} > R_c
 \end{array} \right .
\end{equation} 
together with a self term:
\begin{equation}\label{eq:selfg}
     U^{\rm self}_i = -\frac{1}{2} q_i^2 g(\alpha,R_c)
\end{equation} 
Table \ref{tab:pairpot} provides 
a summary of different pairwise  methods following the scheme above
(Ref.\onlinecite{fukuda13} also discusses the relation amongst
different pairwise potential schemes).
Recall that for the sake of approximating the full 
Coulomb sum, \Eq{infsum}, the Fourier contribution \Eq{fourier}
still needs to be evaluated. Under favorable cases, that may be
achieved by neglecting all of the reciprocal space sum, and
accounting only for the remaining self contribution,
 \Eq{selfew}). 

However, the theoretical
approach explained here shows that in practice all such methods
may be also employed to accelerate the convergence of the real
space summation alone, whether one opts to ignore the
reciprocal space sum or not. Compared to other methodologies,
however, our results have
several advantages:
1) Both the potential and the force remain continuous at $r=R_c$;
2) The force and potential are fully consistent with each other, and
are obtained in a straight forward manner 
(c.f. Ref. \onlinecite{wolf99,zahn02}) 3) The continuity
of the expressions is not merely plugged in for numerical convenience,
but results from a clear and well understood electrodynamic treatment
(c.f. Ref. \onlinecite{fennell06}). 4) The self term is provided and
follows systematically from the theoretical treatment 
(c.f. Ref. \onlinecite{fennell06}).

\begin{table}
\begin{tabular}{c|c|c|c}
\hline
 Pair potential & $f(r;\alpha,R_c)$ & $g(\alpha,R_c)$ &  Limit $\alpha\to 0$  \\
\hline
$u_{\rm Ew}$ & $\phi(r)$ & - & Coulomb  \\
$u_{\rm Wolf}$ & $\phi(r)-\phi(R_c)$ & $\phi(R_c)$ &  Shifted Coulomb \\
$u_{\rm WFG}$ & $\phi(r)-\phi(R_c) - \phi'(R_c) (r-R_c)$ &  $\phi(R_c) -  \phi'(R_c) R_c$ & Force shifted Coulomb \\
$u_{\rm DRF}$ & $\phi(r)-\phi(R_c) - \frac{1}{2}\phi'(R_c)\frac{r^2-R_c^2}{R_c}$ &
$\phi(R_c) -  \frac{1}{2}\phi'(R_c) R_c $ & Reaction Field \\
\hline
\end{tabular}
\caption{\label{tab:pairpot} List of pair potentials and self terms used to approximate
the real space Ewald sum, \protect{\Eq{erfcsum}}, using the general scheme of
\protect{\Eq{pairwisef}}-\protect{\Eq{selfg}}, with
$\phi(r) = \erfc{\alpha r}/r$. The last column indicates the relation of
each model to the Coulomb potential in the limit where $\alpha\to 0$.
The entry $g(\alpha,R_c)$ for $u_{\rm WFG}$ is the self term for the
Wolf-Fennell-Gezelter method, not provided in the original reference.\cite{fennell06}
For the case where \Eq{pairwisef}-\ref{eq:selfg} are employed to approximate the full Coulomb
sum, \Eq{infsum}, an extra self term, $2\frac{\alpha}{\sqrt{\pi}}$ accounting approximately for the Fourier contribution
\Eq{fourier} must be added to $g(\alpha,R_c)$.
}
\end{table}

\section{Results}

In the following section, we will test the ability of different
real--space lattice summations as a means to approximate the full electrostatic
sum of \Eq{infsum}. Whereas such methods are expected to provide a good
convergence for dense fluids of dipolar molecules,\cite{fennell06} we have
chosen to study crystalline and molten ionic salts. This should provide
a more  stringent test than merely polar fluids and therefore allow us to obtain
conclusions of more general validity.

As a test of an ordered ionic solid, we will consider crystalline sodium
chloride. Accordingly, we will assume 
an ordered arrangement of + and - unit charges with lattice spacing $a$,
as in the crystal rock salt lattice.\footnote{This is an
interpenetrated face centered cubic lattice. Plus charges occupy 
positions (0,0,0),(0,1/2,1/2),(1/2,0,1/2),(1/2,1/2,0), while minus
charges occupy analogous positions that are displaced by (1/2,0,0).}
For an ionic molten compound, we again consider configurations of + and
- unit charges thermally sampled from a screened Yukawa fluid with hard sphere
diameter $\sigma$.

The convergence of the average Coulomb energy felt by an ion
is monitored by calculating:
\begin{equation}
  U_i^{\rm R}(R_c) = \sum_{j}^* u(r_{ij})
           + U^{\rm self}_i
\end{equation} 
with $u(r)$ and $ U^{\rm self}_i$ as given by \Eq{pairwisef}-\ref{eq:selfg}, and
the corresponding choice of $f(r;\alpha,R_c)$ and $g(\alpha,R_c)$ as indicated
in Table \ref{tab:pairpot} for different approximate schemes.

\subsection{Charge neutralization schemes with no damping}

In order to make a transparent comparison between DRF and Wolf's
charge neutralization schemes, let us first consider the case
of zero damping, $\alpha=0$ so that $\phi$ adopts
the bare Coulomb form $1/r$. In this case, Wolf's charge neutralization scheme
becomes a mere shifted Coulomb potential, while DRF becomes the 
Reaction Field Method.

\begin{figure}
    \begin{center}
     \caption{Electrostatic potential energy of the NaCl lattice calculated as the direct pairwise summation and by means of Wolf's and DRF methods.}
     \label{EvsRc_NaCl_alpha0}
     \includegraphics[width=0.40\textwidth,angle=-90]{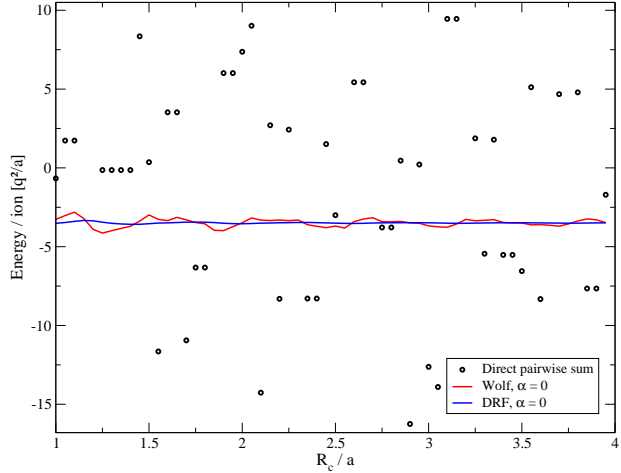}
    \end{center}
 \end{figure}

\begin{figure}
    \begin{center}
     \caption{Electrostatic potential energy of a molten ionic compound of stoichiometry 1:1 calculated as the direct pairwise summation and by means of Wolf's and DRF methods.}
     \label{EvsRc_fundido_alpha0}
     \includegraphics[width=0.40\textwidth,angle=-90]{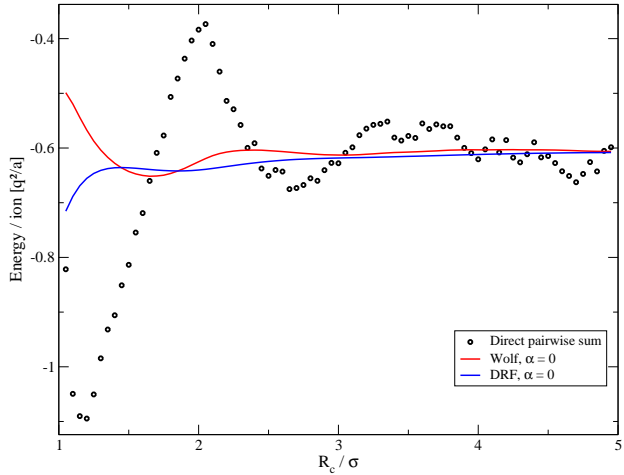}
    \end{center}
 \end{figure}

Figures \ref{EvsRc_NaCl_alpha0} and \ref{EvsRc_fundido_alpha0} 
show $U_i(R_c)$ as estimated by bare Coulomb summation, Wolf's method
and DRF for solid NaCl and the 1:1 molten salt.

As expected, in the NaCl lattice (Fig. \ref{EvsRc_NaCl_alpha0}), the direct summation is not convergent,
and the data is so scattered that it is not even possible to guess approximately  the
exact energy. The introduction of the neutralizing shell proposed by the Wolf's method leads the energy to
converge, and the amplitude of the oscillations decreases remarkably, even
though the damping parameter is zero.
However, if the DRF method is used instead, the amplitude of the oscillations
decreases much more, to the point that they are hardly visible in the scale of
the figure (at least beyond $R_c\approx 1.5 a$).  We have checked that this behavior
also occurs for other  crystalline structures, such as blende and CsCl.
\\
The improved performance of charge neutralization schemes  is also seen in the case of a molten 
ionic compound (Fig. \ref{EvsRc_fundido_alpha0}). The only
difference here is that the direct summation does converge, since, due to the lack of
long-range order, the charges are effectively shielded. The convergence is,
however, very slow. Again, the Wolf's and DRF methods accelerate significantly the convergence, and DRF gives 
rise to somewhat smaller oscillations than Wolf's method.

\subsection{Role of damping}

The convergence of $U_i(R_c)$ may be much improved by using a damping
function $\phi_{\rm mod}(r)=\erfc{\alpha r}/r$. As noted above,
 choosing $\alpha=0$ transforms $\phi_{\rm mod}$ into the bare Coulomb potential.
For finite $\alpha$,  $\phi_{\rm mod}$ becomes a damped Coulomb potential with
a decay rate that is governed by $\alpha$. 

It is important to notice, however, that once $\alpha\ne 0$, then immediately
the Fourier contribution to the Coulomb sum becomes finite, so that 
even in the limit $R_c\to\infty$, the sum $E_i(R_c)$ is  only an
approximation for the full Coulomb sum.

As noted by Wolf, however, there is a wide choice of finite $\alpha$
where the only significant contribution to the Fourier sum, \Eq{fourier}, is the
self term, \Eq{selfew}.
For the NaCl crystal, for example, choosing $\alpha=1.0~a^{-1}$ 
the reciprocal space sum is $\approx 5\cdot 10^{-12}$ 
times smaller than the self term. The case of NaCl is however a
particularly favorable one, and might not be always taken for granted.
For another  simple crystal structure such as CsCl,\footnote{Two interpenetrated
cubic primitive lattices, with plus charges occupying (0,0,0) and minus
charges occupying (1/2,1/2,1/2).}
for example,
the reciprocal space sum is now only  $\approx 10^{-4}$  times smaller
than the self term. A study of the  structure factors
reveals that CsCl is a very unfavorable case because here $\rho(\kvec{}{})$
vanishes only once every two $\kvec{}{}$ vectors, while NaCl is
probably particularly favorable, since $\rho(\kvec{}{})$ vanishes 
for all but about n+1 every $(2n+1)^3$ $\kvec{}{}$ vectors. 

For a fluid, the relative contribution of the sum cannot generally be 
determined, but one here expects that $\rho(\kvec{}{})$ is of finite
range, and should therefore decay faster than for the case of CsCl.

Be as it may, one expects there to be a range of sufficiently 
small $\alpha$, where, on the one hand, the Fourier contribution
is given essentially by the self term, while simultaneously,
the convergence of the real sum is much improved.
One then hopes that the reciprocal space sum can be ignored 
altogether, leaving all of the energy contribution in terms of
a {\em relatively} fast decaying damped coulomb potential.

We test this hypothesis first for a perfect NaCl crystal, then
for the 1:1 molten salt.

\subsubsection{Perfect NaCl crystal} 

\begin{figure}
    \begin{center}
     \caption{Evolution of the real part of the energy with the cut-off radius.
NaCl perfect crystal. $\alpha = 1.0~a^{-1}$.}
     \label{EvsRc_nacl_alpha1.0}
     \includegraphics[width=0.40\textwidth,angle=-90]{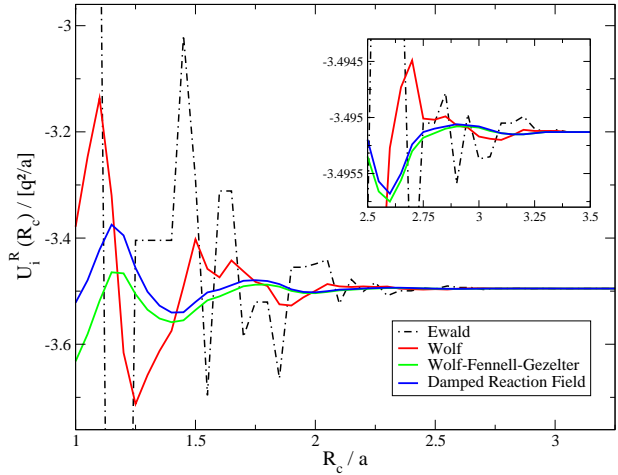}
    \end{center}
 \end{figure}

\begin{figure}
    \begin{center}
     \caption{Evolution of the real part of the energy with the cut-off radius.
NaCl perfect crystal. $\alpha = 2.0~a^{-1}$.}
     \label{EvsRc_nacl_alpha2.0}
     \includegraphics[width=0.40\textwidth,angle=-90]{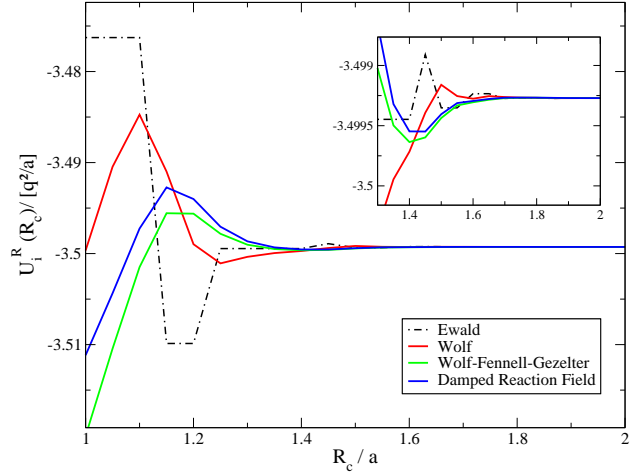}
    \end{center}
 \end{figure}

Figures \ref{EvsRc_nacl_alpha1.0} and \ref{EvsRc_nacl_alpha2.0} show the evolution of the real part of the energy
of the NaCl crystal with the cut-off radius for finite $\alpha$ (1.0~$a^{-1}$
and 2.0~$a^{-1}$ respectively). Results are displayed for the real
space Ewald summation (RSE), Wolf's method (WM), Wolf-Fennell-Gezelter's method (WFG) and DRF.

Notice that for finite $\alpha$, even the real part of the Ewald summation is convergent.
However, the oscillations are rather large and exhibit discontinuities, while
the convergence remains relatively slow. The use of Wolf's method makes the
oscillations decrease strongly. The DRF method shows a clear diminution of the oscillations of the energy, not
only compared with RSE, but also with Wolf's method, while the
performance of WFG is about the same as that afforded by DRF. 
Further increasing $\alpha$ from $1.0~a^{-1}$ to $2.0~a^{-1}$ produces a 
spectacular improvement on the convergence of $U_i(R_c)$. This is obvious right
away by comparing both the $y$ and $x$ scales of figures
\ref{EvsRc_nacl_alpha1.0} and \ref{EvsRc_nacl_alpha2.0}. i.e., not only the
asymptotic value of $U_i(R_c)$ is reached much faster, but also the amplitude of
the oscillations is decreased by more than an order of magnitude. Obviously,
this is at the expense of making the Fourier contribution larger,
and more importantly, increasing the relevance of the reciprocal space sum.
In fact, for $\alpha=2.0~a^{-1}$ the reciprocal space sum
is now already about $10^{-3}$ times the self term of \Eq{selfew}, so
that it becomes unsafe to approximate the Coulomb sum without taking
account of the full Fourier contribution.

An interesting
feature of both WFG and DRF is that, not only is the amplitude of the
oscillations smaller, but also the discontinuities that were apparent in RSE and
to a smaller extent in WM seem to be considerably smoothed. This smoothing
property must be related to the presence of a shifted force contribution to the
effective pair potential, since only WFG and DRF share this feature among the
four methods tested. In fact,  WFG and DRF perform similarly, so that
it would seem it is the the shifted force contribution what makes these
methods perform better than RSE and WM.

At this point it is convenient to mention the significance of the self term,
\Eq{selfg}. Indeed, at first thought one might consider that a constant 
contribution $U^{\rm self}_i$ merely shifts the total energy, but not
the underlying dynamics so that it may be completely ignored.
However, in order to compare the results of $U_i$ for  different methods it
is essential to account for the appropriate self contribution
as indicated in Table \ref{tab:pairpot}. Actually, even for a given
method, results with different $R_c$ (or $\alpha$) can only be compared
when $U^{\rm self}_i$ is included. This is particularly relevant
for the case of WFG, which does not include a self term in the original
reference, and could only be compared with the RSE, WM and DRF by
including the self term of Tab.\ref{tab:pairpot}. This can also be
of great importance in the simulation of open systems, as is the
case of studies performed in the grand canonical ensemble.

As a final remark, we note the conclusions drawn from the analysis performed on
NaCl also hold for other simple crystal structures such as CsCl and ZnS (not
shown).

\subsubsection{Molten salt}

Let us now compare the performance of the pairwise summation schemes for the
1:1 molten salt.

\begin{figure}
    \begin{center}
     \caption{Evolution of the real part of the energy with the cut-off radius.
Molten ionic compound of stoichiometry 1:1. $\alpha = 0.4~\sigma^{-1}$.}
     \label{EvsRc_fundido_alpha0.4}
     \includegraphics[width=0.40\textwidth,angle=-90]{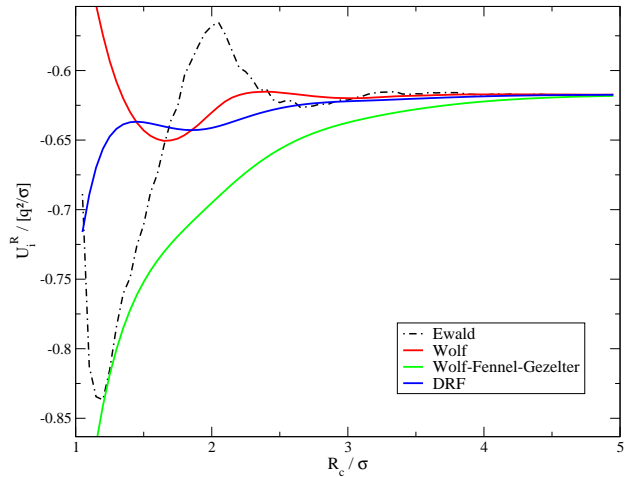}
    \end{center}
 \end{figure}

\begin{figure}
    \begin{center}
     \caption{Evolution of the real part of the energy with the cut-off radius. Molten ionic compound of stoichiometry 1:1. $\alpha = 1.0~\sigma^{-1}$.}
     \label{EvsRc_fundido_alpha1.0}
     \includegraphics[width=0.40\textwidth,angle=-90]{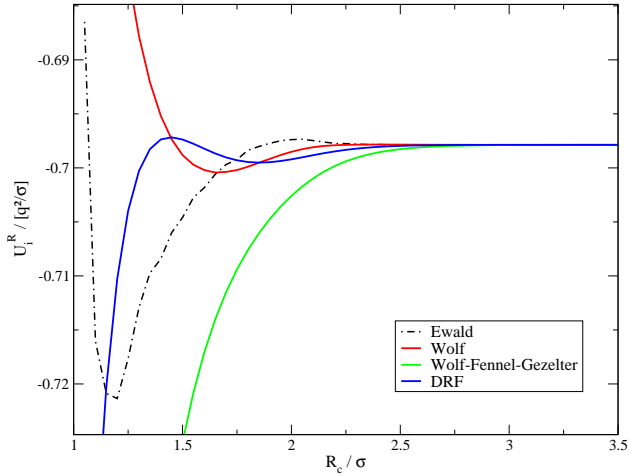}
    \end{center}
 \end{figure}

Depicted in Figures \ref{EvsRc_fundido_alpha0.4} and
\ref{EvsRc_fundido_alpha1.0} is the behavior of $U_i(R_c)$ 
 now calculated for a molten ionic compound. In a molten system, the oscillations of the energy are
smoother than in a crystal, irrespective of the method used. Since the system is
disordered, the interaction with new ions as the cut-off radius increases does
not take place at discrete  distances, but rather, continuously.  \\
Nevertheless, the oscillations given by RSE are still too high, and the
alternative pairwise schemes clearly improve this situation.
In the figures, it is clear that  Wolf's method already diminishes widely the oscillations, and DRF makes this
improvement somewhat more significant. \\
On the contrary, the WFG method now seems to converge slower, and more
importantly, does not seem to oscillate but rather approach the asymptotic
energy from below.

\subsection{Calculation of the virial}

In the preceding sections we have studied the convergence of the energy. Another
important issue refers to the convergence of the forces. Since, however,
the net force exerted on an ion in a perfect crystal is zero, we rather
consider the virial per ion, which has similar convergence issues as the
energy. Accordingly, we study the convergence
of a virial function, defined as:
\begin{equation}
  W_i^R(R_c) = \sum_{j}^* w(r_{ij})
\end{equation}
where
\begin{equation}
 w_{ij} = \overrightarrow{f_{ij}} \cdot \overrightarrow{r_{ij}} 
\end{equation}
while the forces are given by differentiation of the corresponding pairwise
potentials (c.f. Tab.\ref{tab:pairpot}).

\begin{figure}
    \begin{center}
     \caption{Evolution of the virial of an ion with the cut-off radius. NaCl
perfect crystal. $\alpha = 1.0~a^{-1}$.}
     \label{virialvsRc_nacl_alpha1.0}
     \includegraphics[width=0.40\textwidth,angle=-90]{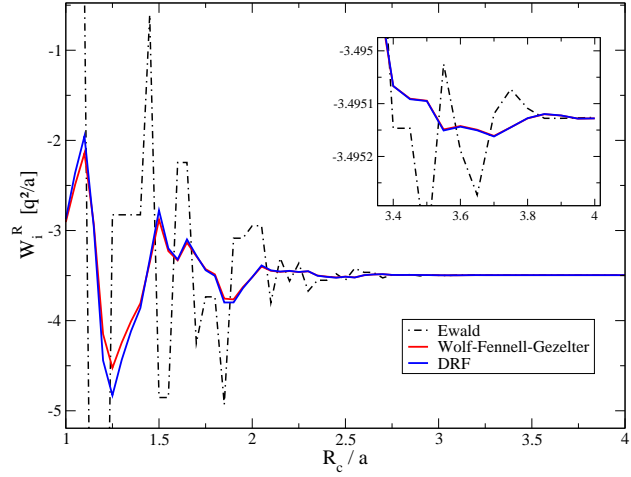}
    \end{center}
 \end{figure}

Figure \ref{virialvsRc_nacl_alpha1.0} shows the evolution
of the virial of an ion inside de NaCl lattice with the cut-off radius
for a fixed value of $\alpha = 1.0$. Results are shown for RSE, WFG, and
DRF. Notice that we do not employ Wolf's method here, as it gives a
discontinuity in the force at $R_c$.

The figure explicitly shows a rather poor convergence of the RSE, 
which exhibits strong oscillations and discontinuities as a function
of $R_c$. At this point, it is worth mentioning that for $\alpha=0$,
the RSE method becomes the bare Coulomb sum and does also not
converge at all (not shown). Using the effective pairwise schemes
very much improves this situation.
Indeed, the amplitude and also the frequency of the oscillations
is reduced, while the convergence is also achieved faster. Apparently,
WFG and DRF perform similarly, lending support to the idea that a
shifted force is necessary (and perhaps sufficient) to improve
the convergence of the sum. This is somewhat confusing, given
that one can hardly attribute any electrostatic significance to
the WFG scheme, which merely corresponds to a force shifted
potential.

\begin{figure}
    \begin{center}
     \caption{Evolution of the total virial with the cut-off radius. Molten
ionic compound with stoichiometry 1:1. $\alpha = 0.4~\sigma^{-1}$.}
     \label{virialtotalvsRc_fundido_alpha0.4}
     \includegraphics[width=0.40\textwidth,angle=-90]{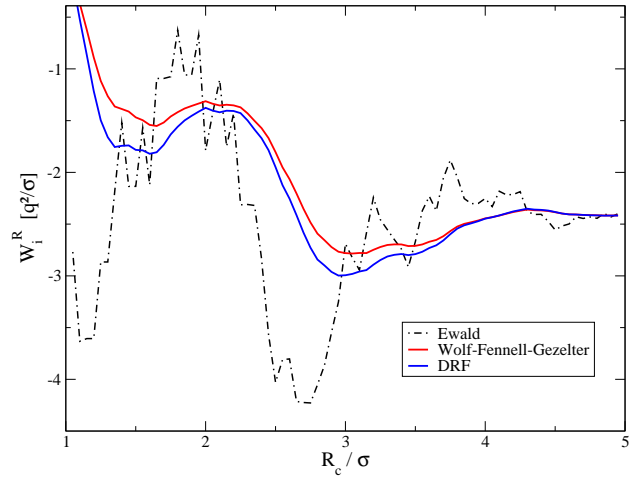}
    \end{center}
 \end{figure}

Figure \ref{virialtotalvsRc_fundido_alpha0.4} displays results
for $W_i^R(R_c)$ in the molten salt system, with $\alpha=0.4~\sigma^{-1}$.
In this case, the direct RSE summation displays deviations of
about the same amplitude than WFG and DRF, but clearly exhibits stronger 
discontinuities. Again both WFG and DRF perform similarly.

\section{Convergence and optimization}

\begin{figure}
    \begin{center}
     \caption{Deviation of the energy from its convergence value (Rc = $\infty$)
as a function of $\alpha$ calculated for several cut-off radii. Comparison
between Ewald summation and DRF method. $\Delta U^F_i(\alpha;n_c=0)$ is 
shown as a black dotted-dashed line in both plots.  NaCl perfect crystal.}
     \label{erorres_partereal_nacl}
     \includegraphics[width=0.40\textwidth,angle=-90]{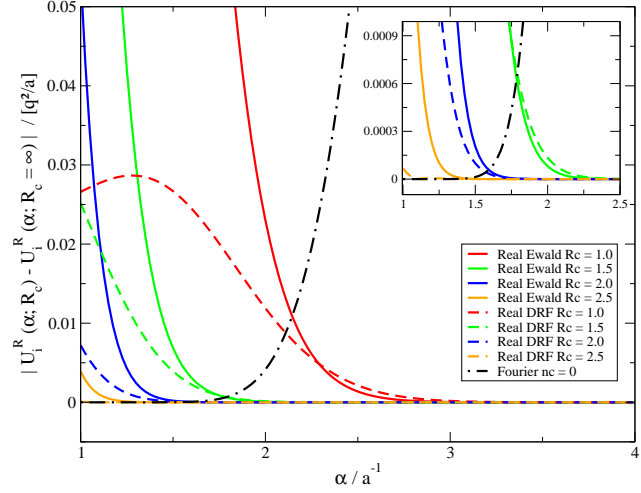}
    \end{center}
 \end{figure}

\begin{figure}
    \begin{center}
     \caption{Deviation of the energy from its convergence value (Rc = $\infty$) as a function of $\alpha$ calculated for several cut-off radii. Comparison between Ewald summation and DRF method.
$\Delta U^F_i(\alpha;n_c=0)$ is 
shown as a black dotted-dashed line in both plots. 
Molten ionic compound with stoichiometry 1:1.}
     \label{erorres_partereal_fundido}
     \includegraphics[width=0.40\textwidth,angle=-90]{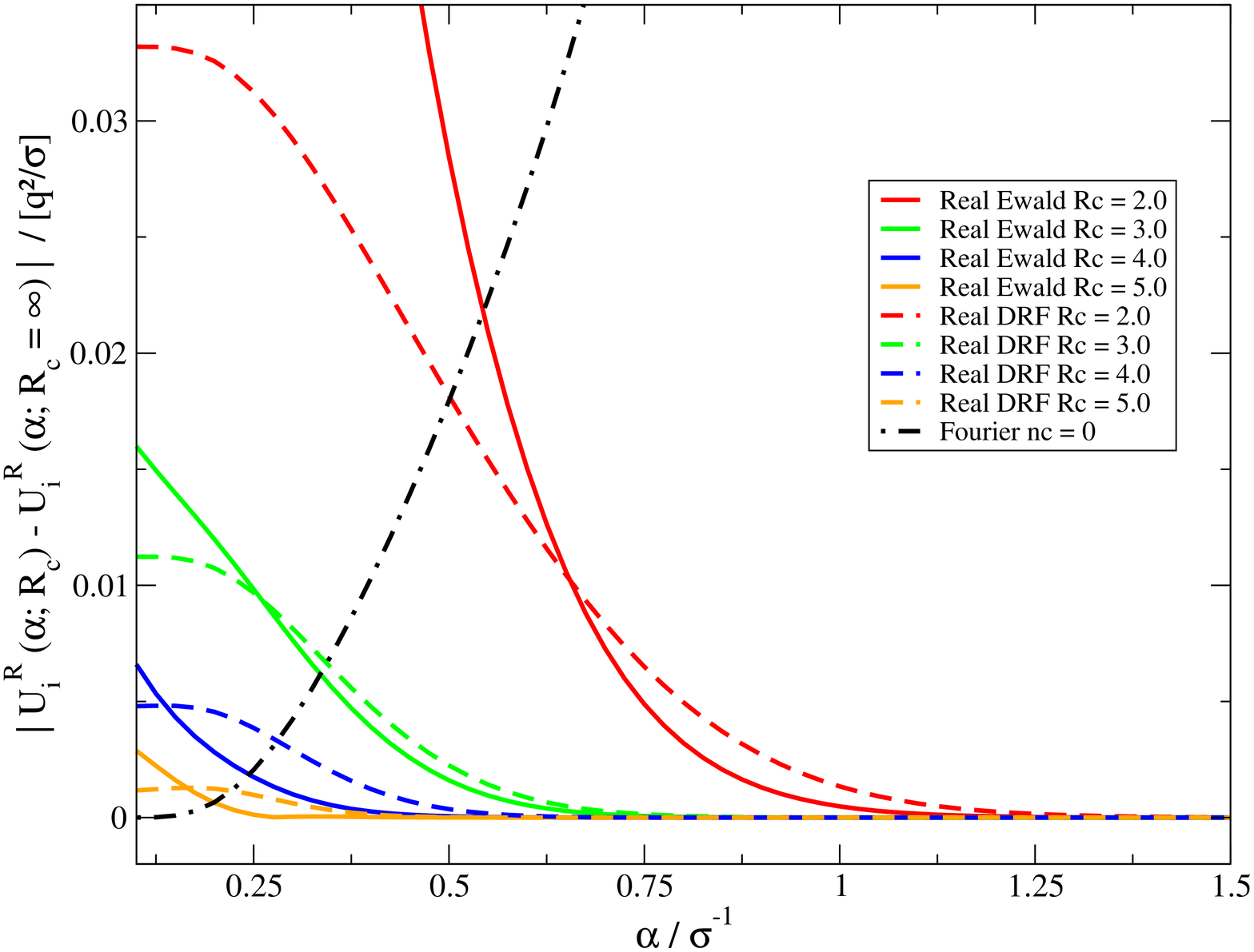}
    \end{center}
 \end{figure}

In the previous sections we have shown analysis similar to those of
Ref.\onlinecite{wolf99,fennell06}, which indicate that for sufficiently
small $\alpha$, one can devise effective pairwise potentials that
allow to approximate the Coulomb sum, \Eq{infsum}, at low computational
cost. Furthermore, we have shown that the damped reaction field method
converges at least as good as  WM and WFG methods, but has a sounder
physical interpretation.

However, our theoretical analysis reveals that DRF, as well as
WM and WFG,  are in fact plausible approximations for the real space
Ewald sum, \Eq{erfcsum}, whether one opts to neglect the reciprocal space sum
(\Eq{fourier}) or not. This is a very relevant issue which has
apparently not been considered previously. It suggest one could perform
the full Ewald summation, accounting both for the real space and Fourier
contributions (whether in the standard implementation or in the particle mesh
approaches \onlinecite{toukmaji96}), but using DRF in order to accelerate 
the convergence
of the real space term. This would allow to perform the Coulomb
sum with arbitrary precision, but decreasing the cost of the real space
sum.

In order to study this question in more depth, we 
introduce a measure of the truncation error performed in the real space sum
as follows:
\begin{equation}\label{eq:rerror}
\Delta U^{\rm R}_i(\alpha ; R_c) = \ \mid U^{\rm R}_i(\alpha ; R_c) - U^{\rm R}_i(\alpha ; R_c = \infty)\mid
\end{equation} 

A plot of the real space convergence error is shown in Fig.\ref{erorres_partereal_nacl} 
for the case of crystalline NaCl. Full lines indicate the error that
results when \Eq{erfcsum} is evaluated by sum of plain $\erfc{\alpha r}/r$
contributions, i.e., RSE, and dashed lines  the case where it is evaluated using DRF.
The figure clearly shows that, for reasonable choices of $R_c$, there is
a range of small $\alpha$, where DRF exhibits much smaller convergence error
than RSE. However, as $\alpha$ increases, the difference becomes smaller,
and actually exhibits a crossover to a regime where the plain RSE seems to perform
better. 

At any rate, the convergence is always improved for large values of  $\alpha$.
However,  one can not increase $\alpha$ arbitrarily, because then the Fourier contribution 
would become relevant. In order to optimize the choice of $\alpha$ for a given
fixed $R_c$, the relevant issue is then what the error of neglecting the Fourier
sum is.

In order to asses this, we now introduce a measure
of the  truncation error performed  in the reciprocal space sum of
\Eq{fourier}. This error results 
from the neglect of  contributions with reciprocal
space vectors ${\bf k}=\frac{2\pi}{L}(n_x,n_y,n_z)$ larger than a prescribed
cutoff, $k_c$. Similarly to \Eq{rerror}, we can therefore 
introduce a Fourier space truncation error as:
\begin{equation}
\Delta U^{F}_i(\alpha ; n_c) = \ \mid U^{F}_i(\alpha ; n_c) - U^{F}_i(\alpha ; n_c = \infty)\mid \\
\end{equation} 
where $n_c$ defines a cubic cutoff for vectors whose integer components
(absolute value) are larger than $n_c$.

Obviously, the pairwise effective potentials as estimated in WM and WFG,
as well as in the zero dipole method of Ref.\cite{fukuda11}, neglect
the Fourier space contribution all together. This corresponds
to assuming a Fourier space convergence error for zero cutoff, $n_c=0$.

Fig.\ref{erorres_partereal_nacl} displays together with
$\Delta U^{\rm R}_i(\alpha ; R_c)$ the Fourier space convergence
error  $\Delta U^{\rm F}_i(\alpha ; n_c)$ for the special case where $n_c=0$.

Clearly, it is always possible to choose a value of $\alpha$ sufficiently small
that $\Delta U^{\rm F}_i(\alpha ; n_c=0)$ is negligible. However,
the optimal choice of $\alpha$ is one where both $\Delta U^{\rm R}_i(\alpha ;
R_c)$  and $\Delta U^{\rm F}_i(\alpha ; n_c)$  are of similar order.
Obviously, it does not make much sense to achieve an exceptional convergence
in the real space contribution, if this is at the cost of having a much
larger reciprocal space error, because what matters is the overall convergence
of the full Coulomb sum.

Unfortunately, Fig.\ref{erorres_partereal_nacl} seems to suggest that $\Delta
U^{\rm F}_i(\alpha ; n_c=0)$
cuts $\Delta U^{\rm R}_i(\alpha ; R_c)$ at a point where there is no significant
advantage of the DRF method over the plain RSE summation.
i.e., the point at which $\Delta U^{\rm F}_i(\alpha ; n_c=0)$ is
about the same order of magnitude of  $\Delta U^{\rm R}_i(\alpha ; R_c)$
occurs where DRF and RSE perform similarly.  This seems
clearly visible, at least for the perfect NaCl crystal and three choices
of $R_c/a=1, 1.5$ and $2$.

The same conclusions may be drawn from Fig.\ref{erorres_partereal_fundido},
which plots the convergence errors for the case of the molten salt.
Here, the situation seems to be even less advantageous for DRF, since
the region of  $\alpha$ where it exhibits smaller error than the plain RSE
is smaller. In fact, in this case $\Delta U^{\rm F}_i(\alpha ; n_c=0)$ 
cuts the curves  $\Delta U^{\rm R}_i(\alpha ; R_c)$ after the crossover
where RSE converges better than DRF, for all but the smallest $R_c$
studied.

We now study the optimization of the Ewald sum by relaxing the
constraint of pairwise schemes, i.e., by allowing for a finite
reciprocal space cutoff. This is still a relevant issue, since
the Ewald sum requires to share
the computational cost between both the real and Fourier
contributions.\cite{kolafa92} An improvement in the convergence
of the real space summation  would allow to shift some of
the cost of the reciprocal space sum to the real space term, making 
the former less expensive.

\begin{figure}
    \begin{center}
     \caption{Quadratic deviation of the real and the reciprocal part of the energy from its convergence value (infinite Rc or nc) as a function of $\alpha$ calculated for several cut-off radii. Comparison between Ewald summation and DRF method. NaCl perfect crystal.}
     \label{erorres_parterealyreciproca_nacl}
     \includegraphics[width=0.40\textwidth,angle=-90]{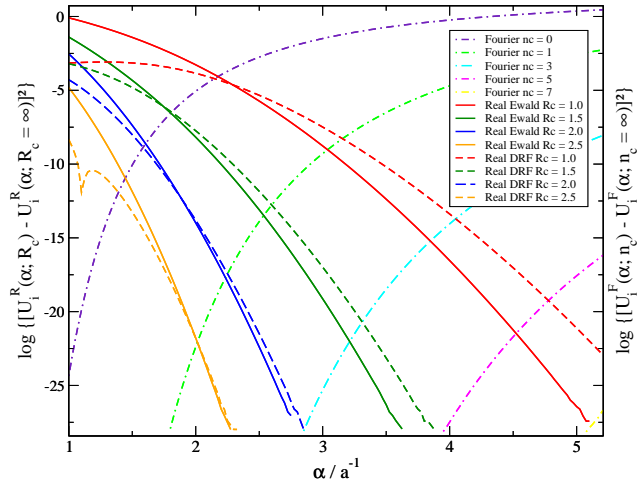}
    \end{center}
 \end{figure}

\begin{figure}
    \begin{center}
     \caption{Quadratic deviation of the real and the reciprocal part of the energy from its convergence value (infinite Rc or nc) as a function of $\alpha$ calculated for several cut-off radii. Comparison between Ewald summation and DRF method. Molten ionic compound with stoichiometry 1:1.}
     \label{erorres_parterealyreciproca_fundido}
     \includegraphics[width=0.40\textwidth,angle=-90]{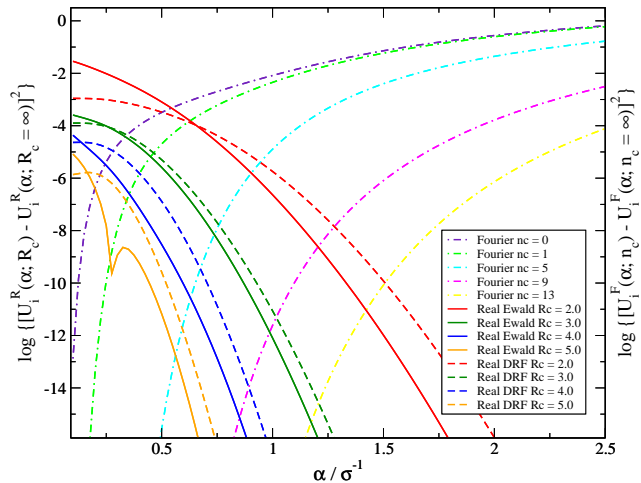}
    \end{center}
\end{figure}

Fig.\ref{erorres_parterealyreciproca_nacl}-\ref{erorres_parterealyreciproca_fundido}
display  $\Delta U^{\rm R}_i(\alpha ; R_c)$ as estimated from RSE and DRF,
compared with $\Delta U^{\rm F}_i(\alpha ; n_c)$ for crystalline and molten
NaCl, respectively. Notice that the errors are shown in logarithmic scale,
with large negative values indicating small errors.

From  inspection of $\Delta U^{\rm R}_i(\alpha ; R_c)$, it now becomes clear
that one can identify two different regimes. Firstly, a {\em reaction field
regime} corresponding to small values of the  product $\alpha R_c$,  where
DRF produces smaller real space errors than the plain Ewald real space sum.
Secondly, an {\em Ewald regime}, of large $\alpha R_c$, where it is actually
the RSE summation which performs better than DRF, and presumably, better
than whichever pairwise effective potential scheme. 

Optimization of the Ewald sum requires to choose $R_c$, $\alpha$ and $n_c$
such that $\Delta U^{\rm R}_i(\alpha ; R_c)\approx \Delta U^{\rm F}_i(\alpha ;
n_c)$. Thus, a mixed scheme using DRF and reciprocal space sums would
be advantageous over the Ewald method if the condition above
is met at a value of $\alpha$ where the convergence error of DRF is
smaller than that of RSE. Unfortunately, our plots seem to indicate
that this is not usually the case. Only for the special choice of $n_c=0$,
i.e., for complete neglect of the reciprocal space sum, we
find that an optimized DRF yields errors similar to the optimized RSE.

Notice that the conclusions drawn above are restricted to
bulk systems. Additional care must be taken for
inhomogeneous systems, where the charge distribution
may exhibit large or even diverging correlations of long
wavelength. The charge structure factor $\rho({\bf k})$ in
\Eq{fourier} then becomes very large for  small ${\bf k}$,
and the reciprocal space sum is significant even
for small $\alpha$. Empirically, this has been observed by 
Takahashi et al., who studied  the liquid-vapor interface
of SPCE/E water and noticed the convergence of Wolf's 
method  could only be achieved for cutoffs as large as
20 molecular diameters.\cite{takahashi11} Such
a poor convergence indicates very long range, small
wave-vector interactions across the interface. These are best dealt
by transferring some of the real space computational burden
to the reciprocal space sum, which precisely is devised
to ensure fast convergence of small wave-vector contributions. 
Theoretically, it has been shown that properly taking
into account a large scale inhomogeneity 
requires to account for the net dipole moment of the system
in the direction perpendicular to the interface. Such
contribution, which is significant and independent of the choice of
$\alpha$, corresponds to a zero wave-vector term of the reciprocal
space sum and cannot possibly be accounted for with a real space
finite cutoff.\cite{deleeuw80,hautman89,spohr97,yeh99}

\section{Conclusions}

In this work we have considered a number of recent methodologies which
allow to approximate the electrostatic energy of charged systems by
means of effective pairwise potentials.\cite{wolf99,fennell06,fukuda11}

Our theoretical study reveals that these methods may be actually considered
as approximations to the real space Ewald sum only, \Eq{erfcsum}, rather
than to the sought Coulomb sum, \Eq{infsum}. 
This is an important matter, because it means that the pairwise effective
potentials cannot possibly account for the long range electrostatic
response of charged systems. Particularly, such methods cannot account ab-initio
for the surface term of the Coulomb energy, which is recovered in the full
Ewald sum as the $k=0$ limit to the reciprocal space
term.\cite{deleeuw80,ballenegger14}

For cases where the damping parameter $\alpha$ is small enough, however,
all terms besides the $k=0$ contribution to the reciprocal space sum are negligible 
and the Ewald self term, \Eq{selfew} is then a very good
approximation for the Fourier space contribution, \Eq{fourier}.

In such cases, our numerical study shows that pairwise effective potentials
produce indeed a better convergence of the real space Ewald sum than the
Ewald pair potential itself. In practice, this may be achieved by introducing
a shifted,\cite{wolf99} or a shifted force,\cite{fennell06} potential. However,
our theoretical analysis shows that the correct electrostatic treatment
may achieve both the potential and force shift by means of a generalization
of the reaction field method,\cite{fukuda11} which damps the Coulomb potential
via an $\erfc{\alpha r}$ function. This damped reaction field method, used
as an approximation to the real space Ewald sum smoothly transforms from
a pure reaction field to  an effective Ewald sum, by tuning the 
damping parameter $\alpha$.

We have studied the performance of the modified Coulomb sums in a particularly
difficult case of an ionic crystal and its melt. This provides a stringent
test of all methods and thus allows to draw more general conclusions. Our
results show that the new damped reaction field method, which has a much
more clear electrostatic significance, is also the method which provides
better convergence of the energy and the virial for both the crystal and the
melt.

A close inspection of the convergence errors allows to identify two
different regimes. Firstly, a reaction field regime, corresponds to
small $\alpha R_c$, where effective pair wise methods such as the
damped reaction field method converge better than the bare $\erfc{\alpha r}/r$
potential. Secondly, an Ewald regime, corresponding to large $\alpha R_c$,
where the bare Ewald potential actually converges better than all
effective pair wise methods studied. Our numerical study shows
that the crossover from the reaction field to the Ewald regime occurs
precisely for those values of $\alpha$ where the reciprocal space sum
has an error about the same size as the real space sum. Since this equality
is the condition for optimization of the Coulomb sum,\cite{kolafa92} the implication is
that for an optimized calculation of the electrostatic energy the Ewald
sum method is not improved by effective pair wise methods.
Such methods do remain useful and produce improved convergence of the
Ewald sum in those cases where high accuracy calculations, or optimization
of the Coulomb sum are not a concern. Alternatively, they might also be
competitive in cases where the availability of parallel computing
does not make competitive the calculation of the reciprocal space sum.

\begin{acknowledgments}
This work was supported by Ministerio de Educacion y Ciencia through project
FIS2010-22047-C05-05.
LGM would like to acknowledge helpful discussions with Sabinne Klapp.
\end{acknowledgments}


%

\end{document}